\documentclass{aa}  
\usepackage{graphicx}
\usepackage{txfonts}
\begin{document}
   \title{On the migration of protoplanets embedded in circumbinary disks}


   \author{A. Pierens
          \and
          R.P Nelson
          }

   \offprints{A. Pierens}

   \institute{Astronomy Unit, Queen Mary, University of London, Mile End Rd, London, E1 4NS, UK\\
     \email{a.pierens@qmul.ac.uk}       
             }

   \date{Received September 15, 1996; accepted March 16, 1997}

 
  \abstract
   {}
   {We present the results of hydrodynamical simulations of low mass
   protoplanets embedded in circumbinary accretion disks. The aim is
   to examine the migration and long term orbital evolution of the protoplanets,
   in order to establish the stability properties of planets that form
   in circumbinary disks.}
   {Simulations were performed using a grid--based hydrodynamics code.
   First we present a set of calculations that study how a binary interacts 
   with a circumbinary disk.  We evolve the system for $\sim 10^5$ binary
   orbits, which is the time needed for the system to reach a
   quasi-equilibrium state. From this time onward the apsidal lines of the
   disk and the binary are aligned, and the binary eccentricity remains
   essentially unchanged with a value of $e_b \sim 0.08$. \\
   Once this stationary state is obtained, we embed a low mass
   protoplanet in the disk and let it evolve under the action of the 
   binary and disk forces. We consider protoplanets
   with masses of $m_p=$ 5, 10 and 20 $M_\oplus$.}
   {In each case, we find that inward migration of the protoplanet is
   stopped at the edge of the tidally truncated cavity formed by the 
   binary. This effect
   is due to positve corotation torques, which can counterbalance the
   net negative Lindblad torques in disk regions where the surface density
   profile has a sufficiently large positive gradient. Halting
   of migration occurs in a region of long--term stability, 
   suggesting that low mass circumbinary planets may be common,
   and that gas giant circumbinary planets should be able to form 
   in circumbinary disks.}
   {}


   \maketitle
%

\section{Introduction}

Among the 215 extrasolar planets discovered to date, about
30 of them reside in binary or multiple star systems (Eggenberger et al. 2004;
Mugrauer, Neuhauser \& Mazeh 2006; Boss 2006).
Most of them are members of stellar binaries
and orbit around one star, on so-called S-type orbits. In
comparison with planets around single stars, the short-period planets
in binary systems
appear to be more massive (Zucker \& Mazeh 2002)
and  tend to a have a very low eccentricity when
their period is  shorter than about 40 days (Eggenberger et al. 2004). \\
Most of the binary systems harbouring planets have a binary
separation $a_b \ge 100 $ AU. However, planets orbiting at $\sim$ 1--2 AU
from one star have also been detected in Gliese 86, $\gamma$ Cephei and HD
41004 A, which are binary systems with $a_b\sim 20$ AU (Eggenberger et al.
2004; Mugrauer \& Neuhauser 2005). For shorter binary separations, a planet 
cannot orbit stably around one star because of the strong 
perturbations due to the binary companion.
The largest stable planetary orbit $a_{max}$ depends on the binary
mass ratio $q_{b}$ and eccentricity $e_{b}$ (Holman \& Wiegert
1999). For a binary with mass ratio $q_{b}=0.3$, 
we have $a_{max}\sim 0.37\; a_b$ for $e_b=0$ and $a_{max}\sim 0.14\;
a_b$ for $e_b=0.5$ (Eggenberger et al. 2004). \\

In principle, close binaries with $ a_b\sim 1$ AU can
harbour planets evolving on a stable P-type orbit which 
encircles the two components of the binary. The stability of such
planets was studied by Holman \& Wiegert (1999), who
provide empirically derived formulae for planetary
stability depending on binary mass ratio and eccentricity.
A circumbinary planet with mass $m_p=2.5\; M_J$
has been detected orbiting at 23 AU from the radio pulsar binary PSR
1620-26. Another with mass $m_p=2.44\; M_J$ has been found to orbit
around the system composed of the star HD 202206 and its 17.4 $M_J$
companion (Udry et al. 2002). Because close binaries are often 
rejected from Doppler radial velocity
surveys, planets have not yet been observed evolving on P-type orbits
around main sequence binary stars.\\
However, several circumbinary disks have been observed around
spectroscopic binaries like DQ Tau, AK Sco and GW Ori. In a few cases 
like GG Tau (Dutrey et al., 1994), the circumbinary disk has been directly
imaged. Observations reveal that this disk
is truncated at its inner edge, which is  a consequence of the tidal torques
exerted by the binary. The truncation radius is 
$\sim 2.7\; a_b$ which is consistent with analytical
estimates from Artymowicz \& Lubow (1994). 
From analytical calculations and numerical simulations, we
expect the truncation radius to range between 
$1.8\; a_b$ and $2.6\; a_b$, depending on the
binary eccentricity (Artymowicz \& Lubow 1994; G{\"u}nther \& Kley 2002;
G{\"u}nther, Sch{\"a}fer \& Kley 2004). \\

Assuming that planets can form inside circumbinary disks, the
fact that such disks have been observed and that 
$\sim 50\; \%$ of the solar-type stars are members of binary or multiple
star systems (Duquennoy \& Mayor 1991) suggests that circumbinary
planets may be common. It is thus of interest to investigate how
planet formation occurs in circumbinary disks. So far, very few
studies have focused on this topic. Moriwaki \& Nakagawa (2004) studied
planetesimal accretion in gas--free circumbinary disks, and found that
planetesimals can grow only in regions farther out than 13 AU from
a binary with $a_b=1$ AU and $e_b=0.1$. The influence of gas may
be important here as it will provide eccentricity damping for the
planetesimals, such that the region in which accretion can occur
lies closer to the binary. Recently, Quintana \& Lissauer (2006)
simulated the late stages of terrestrial planet formation for
$0.05\; AU\le a_b\le 0.4\; AU$ and $0<e_b<0.8$. They found that planetary
systems with properties similar to those around single stars can 
be formed around binaries with $a_b(1+e_b)\le0.2$ AU. Larger values of
$a_b$ and/or $e_b$ lead to sparser planetary systems. \\
Nelson (2003) studied the orbital evolution of a giant planet embedded
in a circumbinary disk. This work indicated that giant
planets  could become trapped in mean motion resonances with the
central binary. This was caused by the planet migrating through the
disk due to type II migration. Subsequent evolution depended on
whether the resonant trapping was stable. Unstable systems suffered
the fate of being ejected by the binary, becoming free-floating
planets while stable systems remained near or at the resonance.\\

Here, we are interested in an earlier stage of
evolution during which the giant
planet is still forming through the growth of a solid core. To address
this issue, we have performed hydrodynamical simulations of low mass
protoplanets evolving in a circumbinary disk.  
We begin by simulating the evolution of a binary--circumbinary disk system.
These simulations are run for $\sim 10^5$
binary orbits until we get a quasi-stationary state, within which
the disk structure and the binary eccentricity remain unchanged.  We
then use this equilibrium state as the initial conditions for the disk and
the binary in subsequent simulations of protoplanets embedded
in circumbinary disks. \\

This paper is organized as follows. In section 2 we describe the
physical model and the numerical setup. In section 3 we present the
results of the simulations. We first describe the results dealing
with binary-disk interactions, and discuss these results in the context
of previous theoretical and numerical work on binary--disk interactions.
We then focus on the evolution of
planets in circumbinary disks. We find that a low-mass protoplanet is
trapped at the edge of the cavity created by the binary and we show that
this effect arises when the corotation and Lindblad torques
cancel each other. We finally discuss our results
in section 4 and present our conclusions.


\section{Equations of motion}
\subsection{Disk evolution}
Assuming that the disk aspect ratio is small, we can consider 
vertically averaged quantities when writing the equations of motion.
The problem is therefore reduced to a two-dimensional one. In polar 
coordinates $(r,\phi)$ and in a frame with the origin located at 
the centre of mass of the binary, the continuity equation reads:
\begin{equation}
\frac{\partial \Sigma}{\partial t}+\nabla\cdot(\Sigma {\bf v})=0
\end{equation}
where $\Sigma = \int^{\infty}_{-\infty} \rho dz$ is the disk surface density.
The equations for the radial and azimuthal components of the disk 
velocity ${\bf v}=(v_r,v_\phi)$ are respectively given by:
\begin{equation}
\frac{\partial \Sigma v_r}{\partial t}+\nabla\cdot(\Sigma v_r{\bf v})-\frac{\Sigma v_\phi^2}{r}=-\frac{\partial p}{\partial r}-\Sigma\frac{\partial \Phi}{\partial r}+f_r
\end{equation}
and
\begin{equation}
\frac{\partial \Sigma v_\phi}{\partial t}+\nabla\cdot(\Sigma v_\phi{\bf v})+\frac{\Sigma v_r v_\phi}{r}=-\frac{1}{r}\frac{\partial p}{\partial \phi}-\frac{\Sigma}{r}\frac{\partial \Phi}{\partial \phi}+f_\phi
\end{equation}
In the above equations, $p$ is the vertically integrated pressure, $f_r$ and $f_\phi$ are respectively the radial and azimuthal components of the vertically averaged viscous force per unit volume. Expressions for $f_r$ and $f_\phi$ can be found for example in Nelson et al. (2000). $\Phi$ is 
the gravitational potential and can be written as:

\begin{equation}
\Phi=\sum_{i=1}^{2}\Phi_{si}+\Phi_p+\Phi_{ind},
\end{equation}
where $\Phi_{si}$ is the potential of the $i^{\rm th}$ 
member of the binary with mass $M_{si}$:
 
\begin{equation}
\Phi_{si}=-\frac{G M_{si}}{|{\bf r}-{\bf r}_{si}|^3},
\end{equation}
and $\Phi_p$ is the potential of the planet with mass $m_p$:

\begin{equation}
\Phi_p = -  \frac{G m_p}{\sqrt{r^2+r_p^2-2rr_p\cos(\phi-\phi_p)+\epsilon^2}}.
\end{equation}
In the previous equation, $\epsilon$ is a softening parameter and 
is chosen to be $\epsilon=0.6 H/r$, where $H/r$ is the disk aspect ratio.
$\Phi_{ind}$ is an indirect term which comes from the fact that the 
frame centered on the binary centre of mass is not inertial. 
This term reads:
\begin{equation}
\Phi_{ind}=G\sum_{i=1}^{2}\frac{M_{si}}{M_*}\left(\frac{m_p({\bf r}_p-{\bf r}_{si})}{|{\bf r}_p-{\bf r}_{si}|^3}+\int_S\frac{dm({\bf r'})({\bf r}'-{\bf r}_{si})}{|{\bf r}'-{\bf r}_{si}|^3}\right).{\bf r},
\end{equation}
where $M_*$ is the total mass of the binary and the integral is 
performed over the surface area of the disk.

\subsection{Binary and planet orbital evolution}
In this work each body can experience the gravitational force due 
to every other one. In other words, we allow the planet to 
gravitationally interact with both the disk and the binary, 
while each member of the binary can interact with the other star and
also with the disk and the planet. The equation of motion for the 
protoplanet is therefore given by:
\begin{equation}
\frac{d^2{\bf r}_p}{dt^2}=-\sum_{i=1}^{2}\frac{GM_{si}({\bf r}_p-{\bf r}_{si})}
{|{\bf r}_p-{\bf r}_{si}|^3}+{\bf f}_{dp}-\nabla \Phi_{ind},
\end{equation}
and the equation of motion for the $i^{\rm th}$ member of the binary is:
\begin{equation}
\frac{d^2{\bf r}_{si}}{dt^2}= -\frac{GM_{sj}({\bf r}_{si}-{\bf r}_{sj})}
{|{\bf r}_{si}-{\bf r}_{sj}|^3}-\frac{Gm_p({\bf r}_{si}-{\bf r}_p)}
{|{\bf r}_p-{\bf r}_{si}|^3}+{\bf f}_{di}-\nabla \Phi_{ind}
\end{equation}
In the previous equations, ${\bf f}_{di}$ is the force due 
to the disk and is defined by:
\begin{equation}
{\bf f}_{di}=\int_S\frac{\Sigma({\bf r}') d{\bf r}'}{\sqrt{r'^2+r_i^2-
2r'r_i\cos(\phi'-\phi_i)+\epsilon^2}}
\end{equation}
Note that if this force is applied to the binary the softening 
parameter $\epsilon$ is set to 0, and if this force is applied to the planet
we exclude the material located inside the Roche lobe of the planet 
$R_H=a_p(m_p/3M_\odot)^{1/3}$. 
Previous simulations have demonstrated that 2D simulations
of protoplanets embedded in disks give migration rates in good agreement
with 3D results if $\epsilon = 0.6H$ (Nelson \& Papaloizou 2004), 
which is why we adopt this value for the softening.
As already mentioned, 
the last term $-\nabla \Phi_{ind}$ arises because the frame 
centrered on the binary centre of mass is not inertial. 

\section{Numerical setup}
\subsection{Numerical method}
The Navier-Stokes equations are solved using the hydrodynamic code GENESIS,
which is basically a 2D ZEUS-like code. It uses a staggered mesh 
and solves the equations of motion for the disk by means of finite 
differences. The numerical method used in this code is spatially second-order
and the implemented advection scheme is based on the monotonic transport
algorithm (Van Leer 1977). Although the FARGO algorithm (Masset
  2000) is implemented in GENESIS, it was not used for the simulations
presented in this paper. The evolution of the planet and binary orbits
are computed using a fifth-order Runge-Kutta integrator (Press et al. 1992). \\
GENESIS  has already been employed in a project aimed at comparing
several hydrocodes on the disk-planet interaction problem 
(de Val-Borro et al., 2006). In this context, the
runs performed with the GENESIS code give similar results to those obtained
using standard hydrodynamics codes such as ZEUS or NIRVANA.\\
For most of the calculations presented here, we use $N_r=256$ grid 
cells in radius and $N_{\phi}=380$ grid cells in azimuth. 
Some low resolution runs of binary-disk interactions using 
$N_r=128$ and $N_\phi=128$ grid cells have also been performed. 
These are long-term evolution calculations, aimed at showing that 
binary-disk interactions lead to an equilibrium state where the disk 
structure, as well as the binary eccentricity remain constant in time. 
This stationary state is obtained after about $5\times 10^5$ binary 
orbits and would require several months of run-time to be reached at 
higher resolution.
\subsection{Computational units}
We adopt computational units in which the total mass of the binary is $M_*=1$, the gravitational constant is $G=1$, and the radius $r=2$ in the computational domain corresponds to 5 AU. The unit of time is $\Omega^{-1}=\sqrt{GM_*/a_b^3}$, where $a_b$ is the initial semi-major axis of the binary which is set to $a_b=0.4$ in this work. This corresponds to an initial separation between the two stars of $\sim 1$ AU. If any planet is present in the disk, it is initially located at $r_p=2.5$, which would correspond to $r_p=6.2$ AU in physical units.

\subsection{Initial and boundary conditions}
In the model that we adopt for the simulations presented here, the 
disk aspect ratio is constant and equal to $H/r=0.05$, 
which is a typical value for protoplanetary disks. The inner edge 
of the computational domain is located at $r_{in}=0.5$ and the outer 
edge at $r_{out}=6$.\\
We use closed, reflecting boundary conditions at both the inner and outer 
radial boundaries. Because we are interested in long-term 
evolution runs, this ensures that the disk mass does not vary. 
There is no evidence of wave reflection at the inner boundary, in spite of 
the use of reflecting boundary conditions. This is because the inner 
boundary is located deep inside the tidally truncated cavity created by 
the binary, and as a consequence, the density there is very small. 
In order to avoid wave reflection/excitation at the outer boundary, 
we impose a low-density region from $r=4$ to $r=6$ using a taper function. \\
From $r=0.5$ to $r=4$ , the initial surface density distribution is
set to $\Sigma(r)=\Sigma_0r^{-1/2}$, where $\Sigma_0$ is defined in
such a way that the disk would contain $0.01$ $M_\odot$ inside 10 AU
(assuming that the mass of the binary is $M_*=1M_\odot$).\\

The anomalous viscosity in the disk, which probably arises from MHD 
turbulence in nature, is modelled using a standard ``alpha" prescription for 
the effective kinematic viscosity $\nu=\alpha c_s H$ 
(Shakura \& Sunyaev 1973), where $c_s$ is the isothermal sound speed. 
Canonical mass accretion rates of $\simeq 10^{-8}$ M$_{\odot}$ per year
observed to occur in T Tauri systems require $\alpha$ values in the
range $10^{-2}$ -- $10^{-3}$. For the purposes of this study, however,
we have adopted a lower value of $\alpha = 10^{-4}$. The reason for this
is that in earlier test simulations, larger values of $\alpha$ caused 
the binary semimajor axis to decrease too rapidly to allow a steady state disk
structure and binary eccentricity to be established. Shrinkage of the binary
orbit eventually causes the important 1:2 and 1:3 Lindblad resonances to
be located interior to the  inner boundary of the numerical grid rather
than within its body, such that the simulations become unrealistic.
The smaller value of $\alpha$ adopted prevents this from occuring
and allows a steady state disk structure to be established (albeit after
very long run times). 

Due to the very long run times required to establish a quasi--steady state
for the binary--circumbinary disk system, we have only been able to
consider a single binary mass ratio $q_{b}=0.1$. The binary is
initiated on a circular, Keplerian orbit with $a_{b}=0.4$ in code units.


\section{Results}
\subsection{Evolution of the binary-disk system}
We begin by focusing on the interaction between the disk and the binary,
and on the joint evolution of these two bodies. The aim of these 
simulations is to demonstrate that the system can reach a quasi-equilibrium 
state in terms of disk structure and orbital elements of the binary. 
This near--stationary state will be used as an initial condition for 
calculations dealing with the evolution of embedded planets in 
circumbinary disks.\\

\subsubsection{Theoretical Expectations}
\label{theory}
Before presenting the simulation results, we discuss the
expected behaviour of the system based on previous numerical
and theoretical work. Early work on close binary systems interacting
with accretion disks focussed on the evolution of disks surrounding
one of the binary components (the primary). Of particular interest
was the superhump phenomenon observed in the so--called SU UMa subclass of
Cataclysmic Variable systems. These systems have fairly extreme
binary mass ratios $q_{b} \simeq 0.1$, and the superhump phenomenon
(characterised as a modulation of the lightcurve evolution)
was interpreted as being due to a precessing, eccentric disk
(e.g. Whitehurst 1988; Hirose \& Osaki 1990; Lubow 1991).
Lubow (1991) presented a fluid dynamical theory that explained the origin of 
the eccentric disk as being due to an instability at the 3:1 Lindblad
resonance generated by non linear mode coupling, 
leading to the growth of disk eccentricity for a circular binary orbit.

The interaction between massive companions (giant planets and brown dwarfs)
and accretion disks in which they were initially embedded was considered
by Papaloizou, Nelson \& Masset (2001). Here the tidal interaction
creates an inner cavity within which the primary star and companion orbit,
and as such these systems are similar to the binary plus circumbinary disks
that we are concerned with in this paper.
Papaloizou et al. (2001) showed that a massive companion on a circular
orbit could cause the surrounding disk to become eccentric. The origin
of the eccentricity was found to be due to a similar instability mechanism 
to that proposed by Lubow (1991), namely non linear coupling between an initial
$m=1$ eccentric disturbance in the disk and the $m=1$ component of
the binary potential leading to an $m=2$ wave being excited at the 1:3
resonance in the circumbinary disk. Based on this we expect to see the 
circumbinary disk in our simulations become eccentric.
Eccentricity growth will saturate when viscous damping matches
the eccentricity forcing rate. Secular interaction
between the eccentric disk and binary should force the binary 
to become eccentric also.

There are important resonant interactions between the disk and binary
that can lead to modification of the binary orbital elements. For a binary
on a modestly eccentric orbit interaction at outer Lindblad resonances
is expected to cause a decay of the semimajor axis in a viscous disk,
where the decay rate depends on the disk viscosity.
Interaction at eccentric Lindblad resonances is expected to cause 
growth of the eccentricity. Working to first order in the binary eccentricity,
$e_{b}$, the 1:2 corotation resonance should induce
eccentricity damping and the 1:3 eccentric Lindblad resonance
should cause eccentricity growth. If the disk is tidally truncated
beyond the 1:2 Lindblad resonance, as is expected for massive companions
(e.g. Artymowicz 1992; Lin \& Papaloizou 1993),
then eccentricity growth is expected.
Thus we expect the eccentricity of the binary system to grow in our
simulations, provided the disk inner cavity does not
extend beyond the 1:3 resonance. Should this occur the eccentricity
should still grow, but at a lower rate since higher order
resonances will be required to drive the eccentricity.
At the present time there is not a well developed theory that can
be used to predict when the binary eccentricity should saturate.
We note, however, that as the eccentricity grows the interaction
at the most significant eccentric Lindblad resonances may become
increasingly non linear such that they saturate (i.e. the density
there is decreased), causing the growth rate to slow.

We see from the above discussion that we expect there to be both
secular and resonant interactions occuring between the disk and binary,
leading to the growth of their eccentricity. When the angular momentum
content of disk and binary are similar then we expect that they
may participate in a joint secular mode in which they precess at
the same rate (Papaloizou 2002). In fact the angular momentum in 
the binary exceeds that in the disk by about a factor of 4
in our simulations, so the existence of a joint mode is probably
marginal.
When their apsidal lines are
misaligned the disk and binary will exert secular torques leading to
changes in their eccentricities (by analogy with the Jupiter--Saturn system).
If the apsidal lines become
closely aligned, however, then these torques will diminish.
Thus, we expect that a steady state configuration will consist
of an eccentric binary system surrounded by an eccentric disk
precessing at the same rate in a prograde direction with
with apsidal lines closely aligned. 

\begin{figure}
   \centering
   \includegraphics[width=\columnwidth]{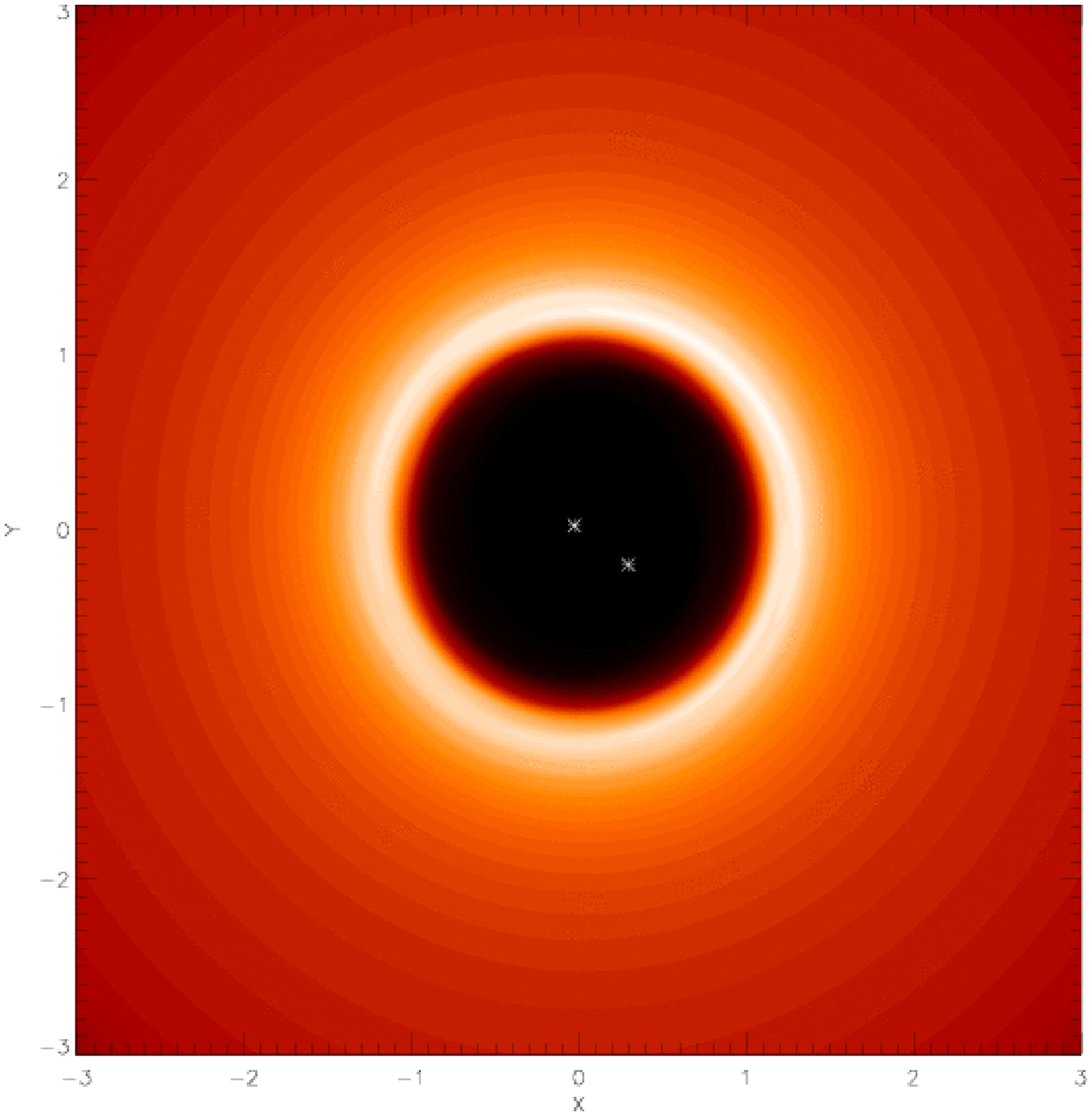}
    \includegraphics[width=\columnwidth]{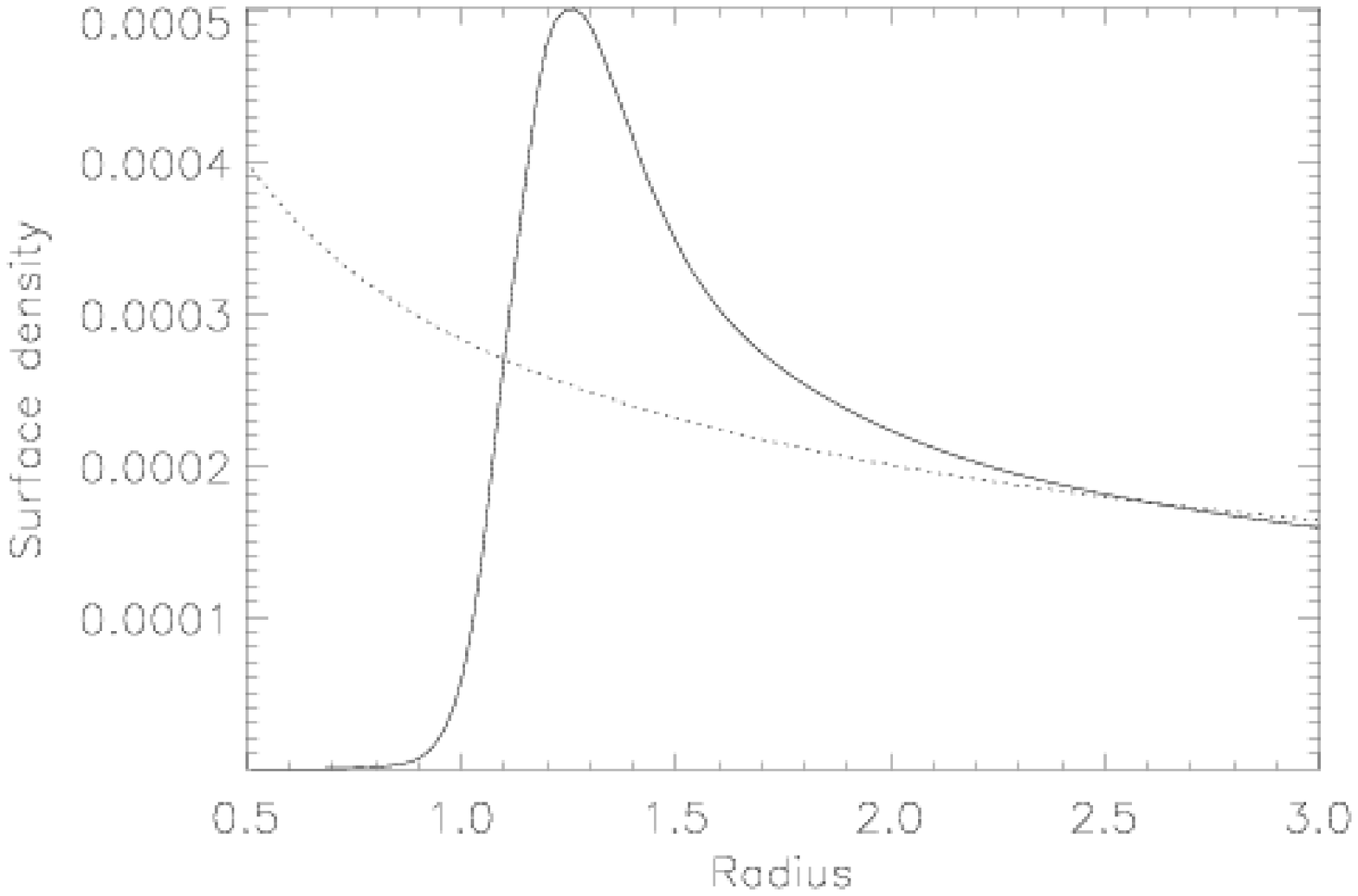}
      \caption{The upper panel shows a snapshot of disk surface density after $1.2\times 10^5$ binary orbits. The lower panel displays the azimuthal average of the surface density at the same time.
              }
         \label{rho_459}
   \end{figure}

\subsubsection{Simulation results}
We now discuss the simulation results for the binary plus circumbinary
disk runs. The upper panel of Figure \ref{rho_459} shows a snapshot of the disk
surface density after $1.2\times 10^5$ binary orbits. Here, the disk
surrounds a binary with mass ratio $q_{b}=0.1$. The binary initially
evolves on a circular orbit and the initial separation between the two
stars is $a_b=0.4$. The lower panel displays the corresponding
azimuthal average of
the disk surface density, as well as the initial
surface density profile. Torques exerted by the binary truncate the
inner edge of the disk. We find that the gap size is $\sim 2.5a$,
which is consistent with analytical estimates from Artymowicz \& Lubow
(1994). \\
 \begin{figure}
   \centering
   \includegraphics[width=\columnwidth,angle=-0]{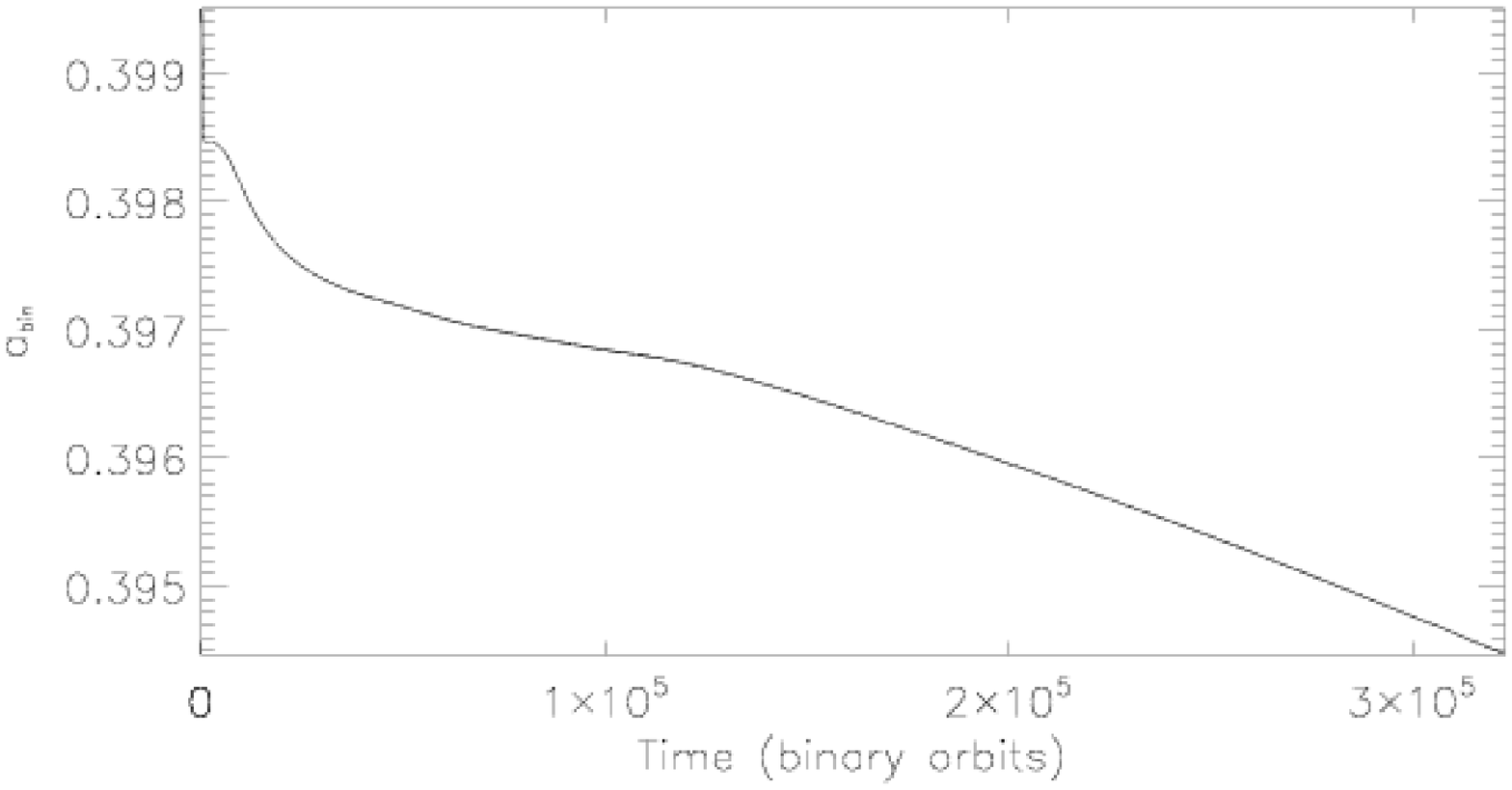}
    \includegraphics[width=\columnwidth,angle=-0]{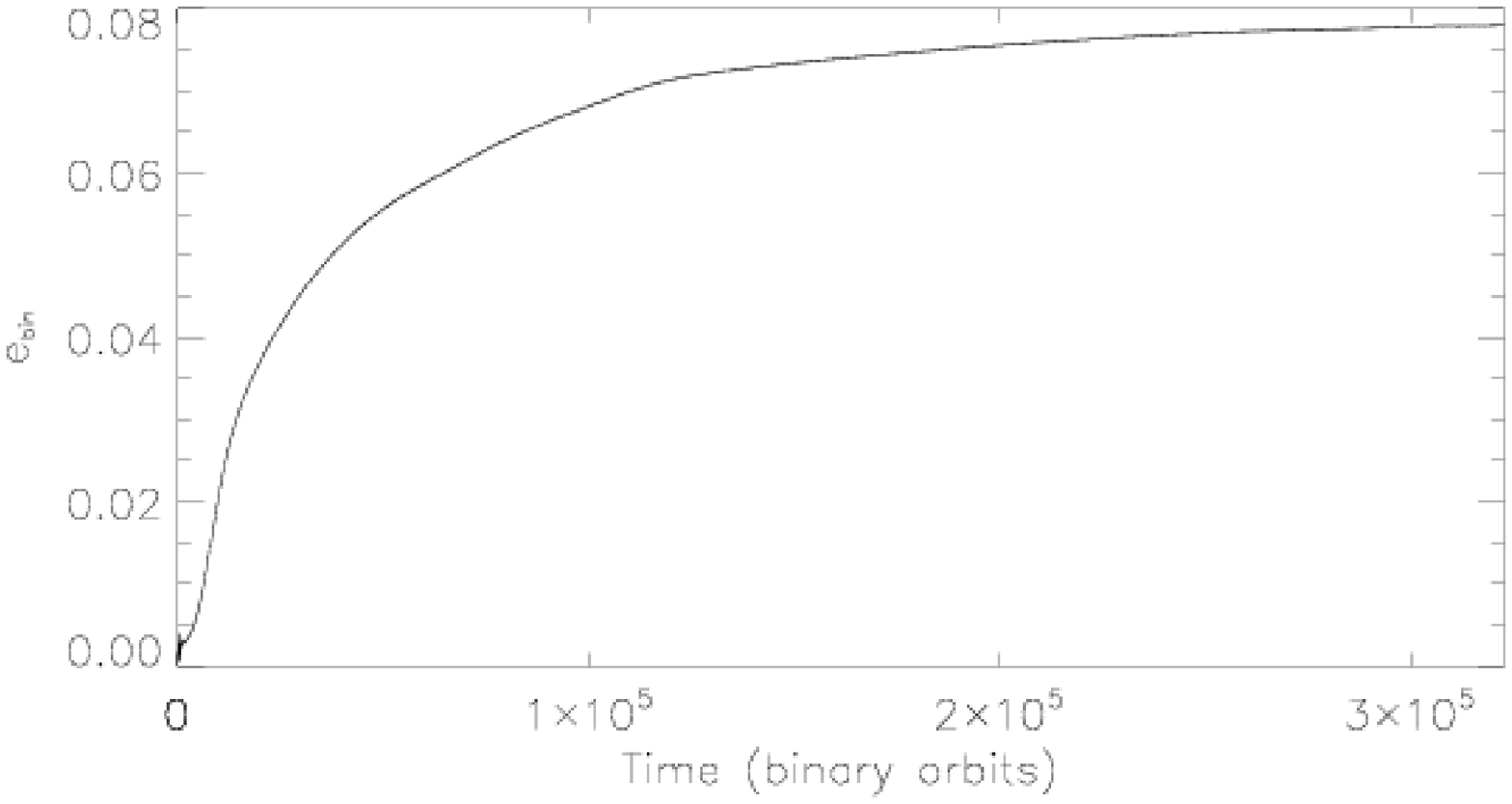}
      \caption{Evolution of the binary semi-major-axis (upper panel) and eccentricity (lower panel) as a function of time, for low-resolution simulations with $128\times 128$ grid cells.
              }
         \label{orbit_bin}
   \end{figure}

Figure \ref{orbit_bin} shows the evolution of the binary semi-major axis and eccentricity as a function of time, deduced from low-resolution simulations with $128\times 128$ grid cells.
As a result of angular momentum being transferred to the disk, 
the binary separation shrinks at a rate $da/dt\sim 10^{-8}$. 
This value is consistent with the orbital decay that we would expect 
from analytical estimates (i.e Armitage \& Natarajan 2005):
\begin{equation}
\frac{da}{dt}\propto -\alpha\left(\frac{h}{r}\right)^2 \frac{M_d}{M_2}\Omega a
\end{equation}
where $M_d$ is the disk mass and $M_2$ the mass of the secondary. The
orbital decay of the binary is slow enough that the 1:3 commensurability
at $r\sim 2.08 \;a_{b}$, which corresponds to the eccentric
Lindblad resonance which is expected to be important in determining the 
evolution of the binary orbit,
always resides inside the computational domain. This makes it
possible to do long-term evolution runs over $\sim 10^5$ binary
orbits, at least at low-resolution. We note that the location of
the 1:3 resonance is just at the base of the gap in fig.~\ref{rho_459},
showing that the density there is quite low. 

The lower panel in fig.~\ref{orbit_bin} shows that the binary 
eccentricity, after some transients lasting for $\sim 5\times 10^3$, 
grows until it saturates at $e \sim 0.08$. The initial 
eccentricity growth is due to the resonant nature of the 
interaction between the disk and the binary. This interaction is expected to
occur mainly at the location of the 1:3 outer Lindblad resonance, 
which promotes eccentricity growth. As time goes by and the 
eccentricity grows, there is evidence that this resonance
saturates due to non linear effects (i.e. the density at the resonance
is decreased), and this saturation combined with higher-order resonances 
coming into play (Artymowicz 1992) causes the eccentricity to reach a 
steady value. We also need to consider the secular interaction
between disk and binary, as this also plays an important role.

\begin{figure}
\centering
 
    \includegraphics[width=\columnwidth,angle=-0]{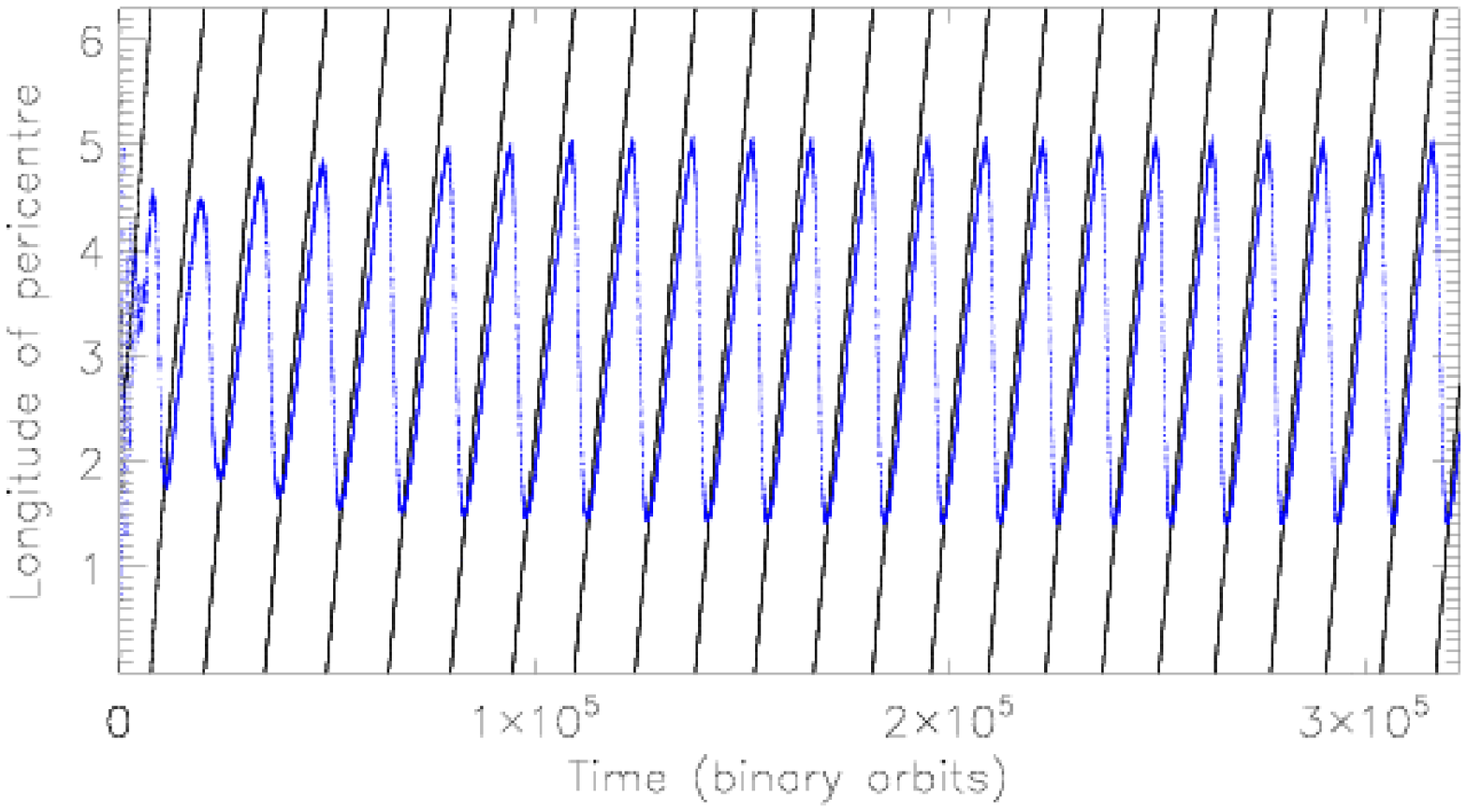}
  \includegraphics[width=\columnwidth,angle=-0]{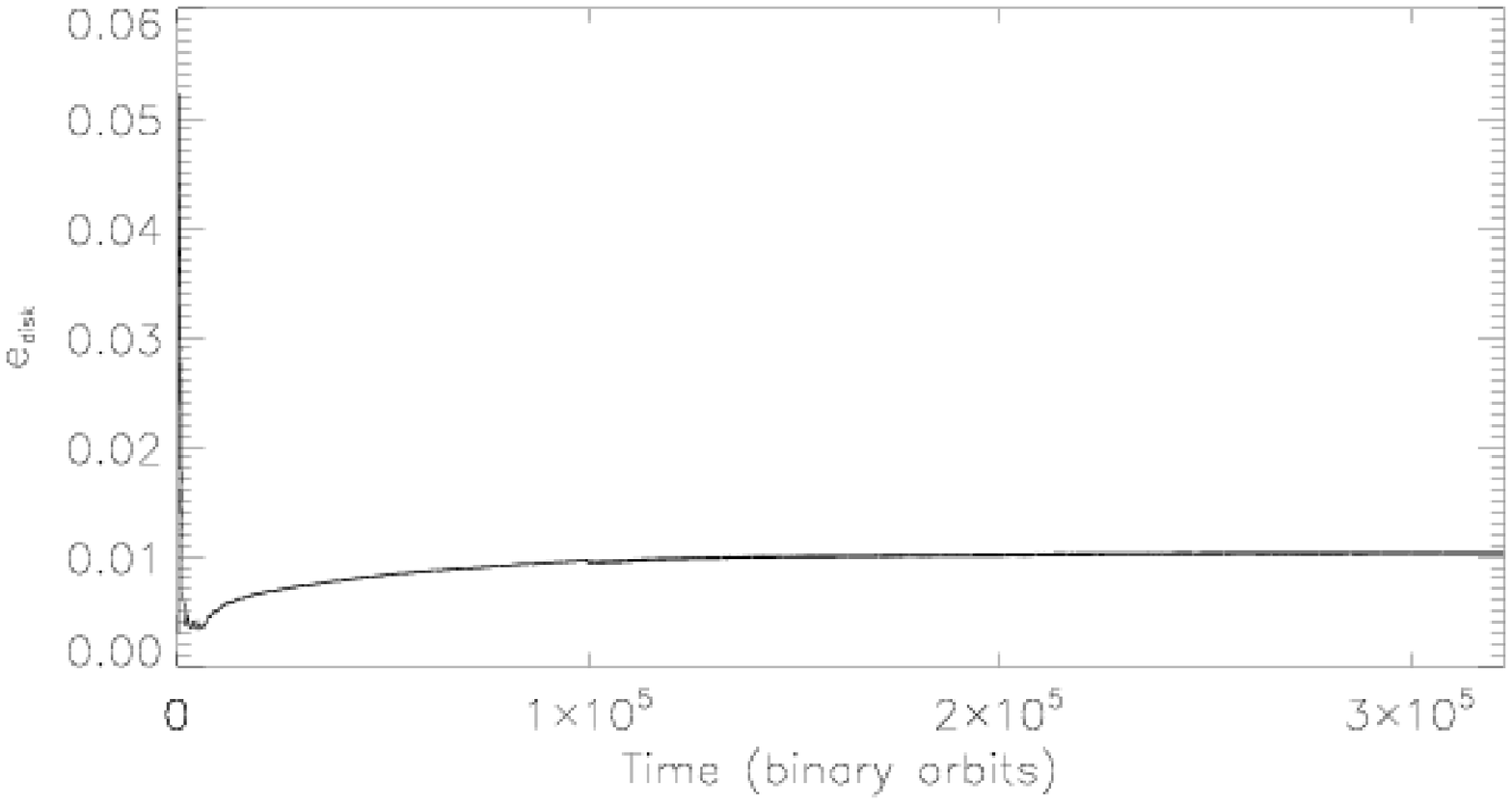}
      \caption{The upper panel shows, as a function time, 
the evolution of the longitudes of pericentre for both the binary 
(black line) and the disk (blue line), deduced from a low-resolution 
simulation with $128\times 128$ grid cells. 
The lower panel displays the evolution of the disk eccentricity.  
              }
         \label{w_disk+bin}
   \end{figure}

We plot the evolution of the longitudes of pericentre for the disk
($\omega_d$) and
binary ($\omega_{b}$) in the upper panel of fig.~\ref{w_disk+bin}.
The disk eccentricity evolution is plotted in the lower panel.
The density--weighted $\omega_d$ is calculated according to the 
definition:
\begin{equation}
\omega_d=\frac{\int_0^{2\pi}\int_{r_{in}}^{r_{max}}\Sigma\; \omega_c dS}{\int_0^{2\pi}\int_{r_{in}}^{r_{max}}\Sigma\; dS}
\end{equation}
where $\omega_c$ is the longitude of pericentre computed inside each
grid cell, and $r_{max}=3$ for the simulations presented here. 
Test calculations have shown that $\omega_d$ does not depend strongly 
on the value of $r_{max}$. 
The figure shows that both the binary and disk precess in a
prograde sense, and that at late times 
the apsidal lines of the disk and binary come into almost
perfect alignment. This alignment coincides with the time when
the binary eccentricity reaches a steady value, 
showing that the secular interaction 
between disk and binary has a significant effect on the eccentricity
evolution of the binary. 

The disk eccentricity, $e_d$, shown in the lower panel of
fig.~\ref{w_disk+bin}, is defined in the simulations by:
\begin{equation}
e_d=\frac{\int_0^{2\pi}\int_{r_{in}}^{r_{max}}\Sigma\; e_c dS}{\int_0^{2\pi}\int_{r_{in}}^{r_{max}}\Sigma\; dS}
\end{equation}
where $e_c$ is the eccentricity computed at the center of each grid cell. 
The disk eccentricity evolves similarly to  the 
binary eccentricity and saturates at the relatively small value of
$e_d\sim 0.01$. 

In line with the expectations outlined in section~\ref{theory},
the disk and binary become eccentric. They achieve a steady state
configuration when the apsidal lines of disk and binary are aligned
and both disk and binary precess at the same rate in a prograde sense.

\begin{figure}
   \centering
   \includegraphics[width=\columnwidth]{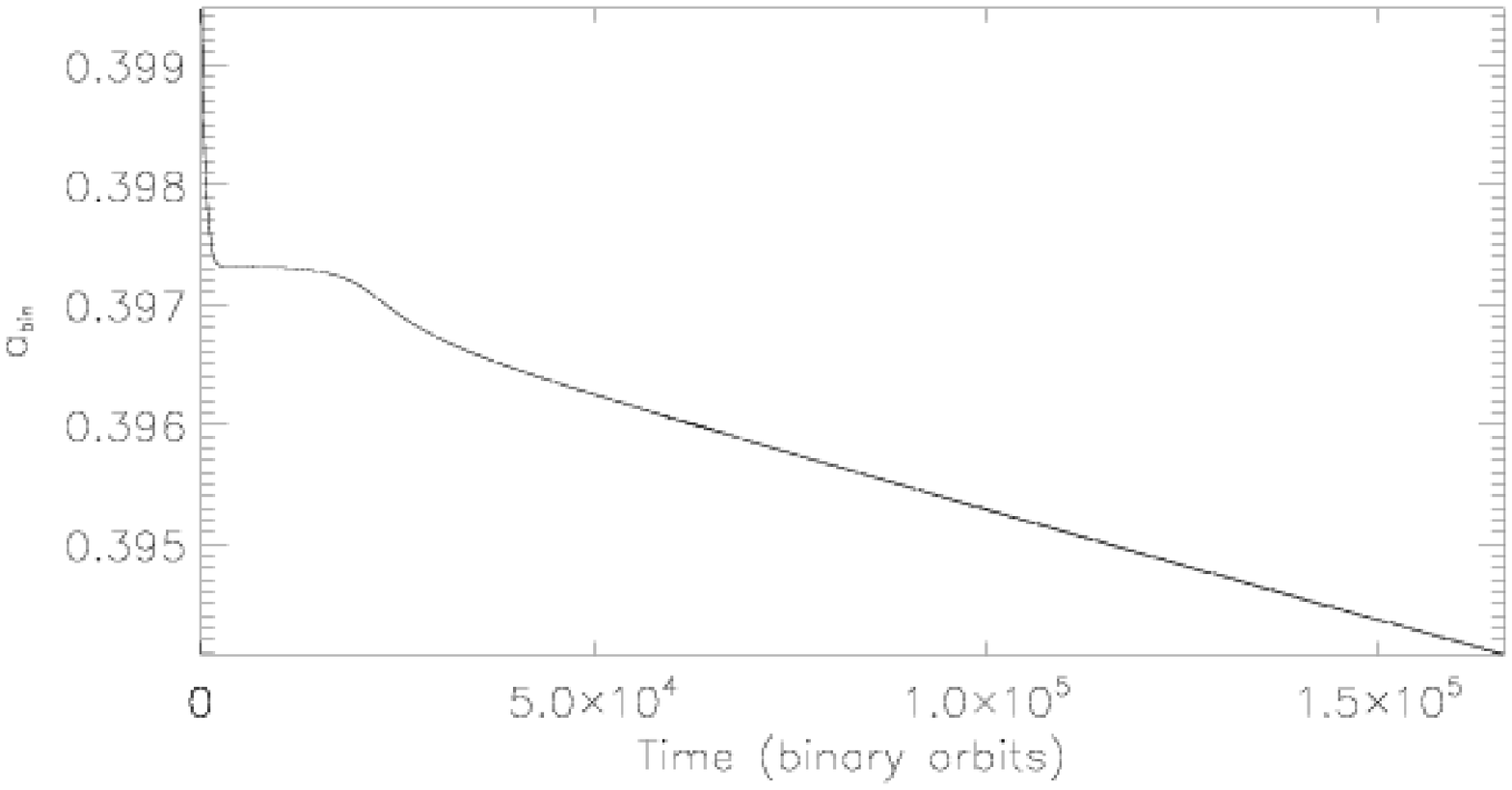}
    \includegraphics[width=\columnwidth]{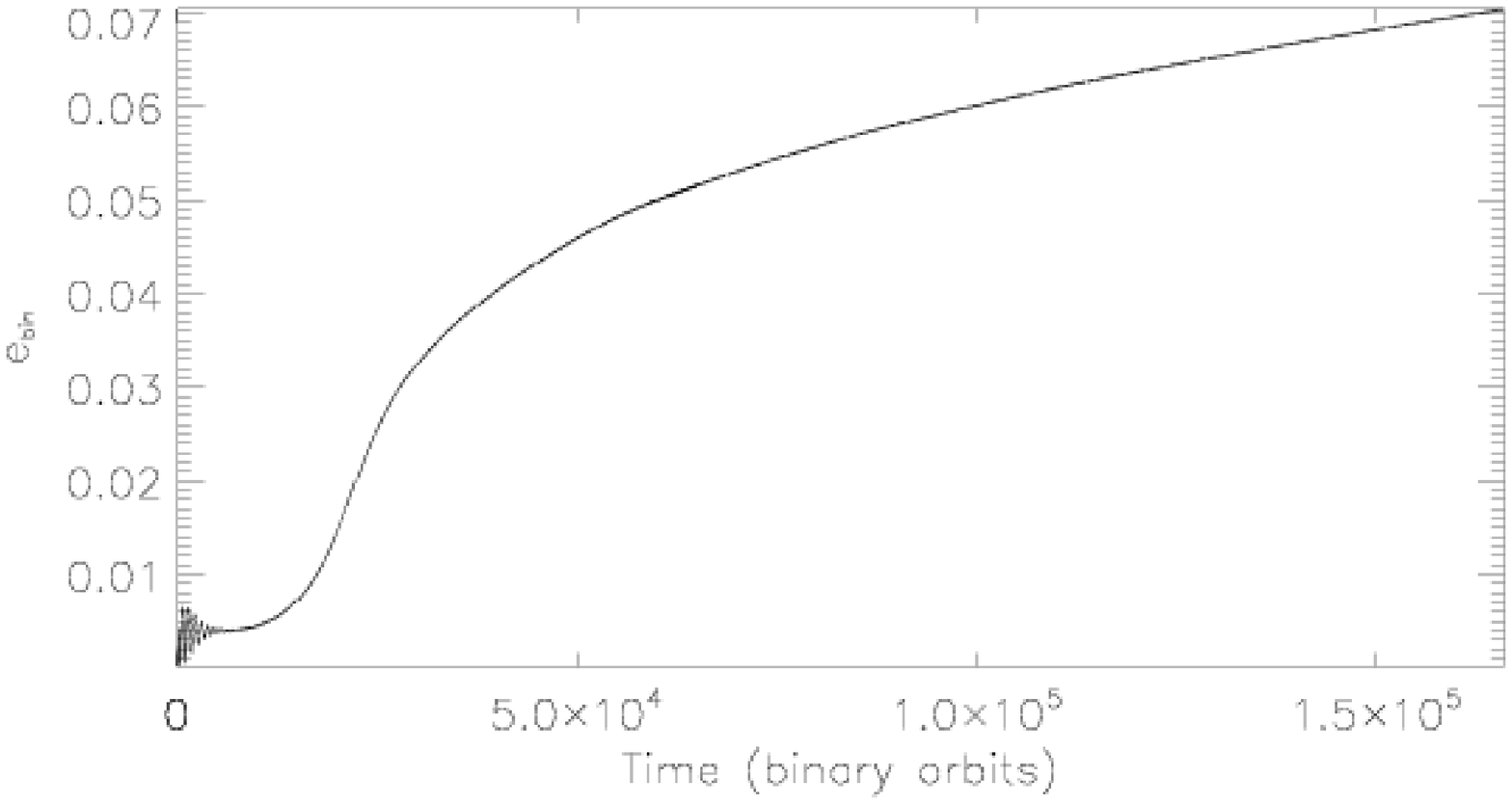}
      \caption{Same as Figure \ref{orbit_bin} but for high-resolution simulations with $256\times 380$ grid cells.
              }
         \label{orbit_bin_hr}
\end{figure}

\begin{figure}
\centering
   \includegraphics[width=\columnwidth]{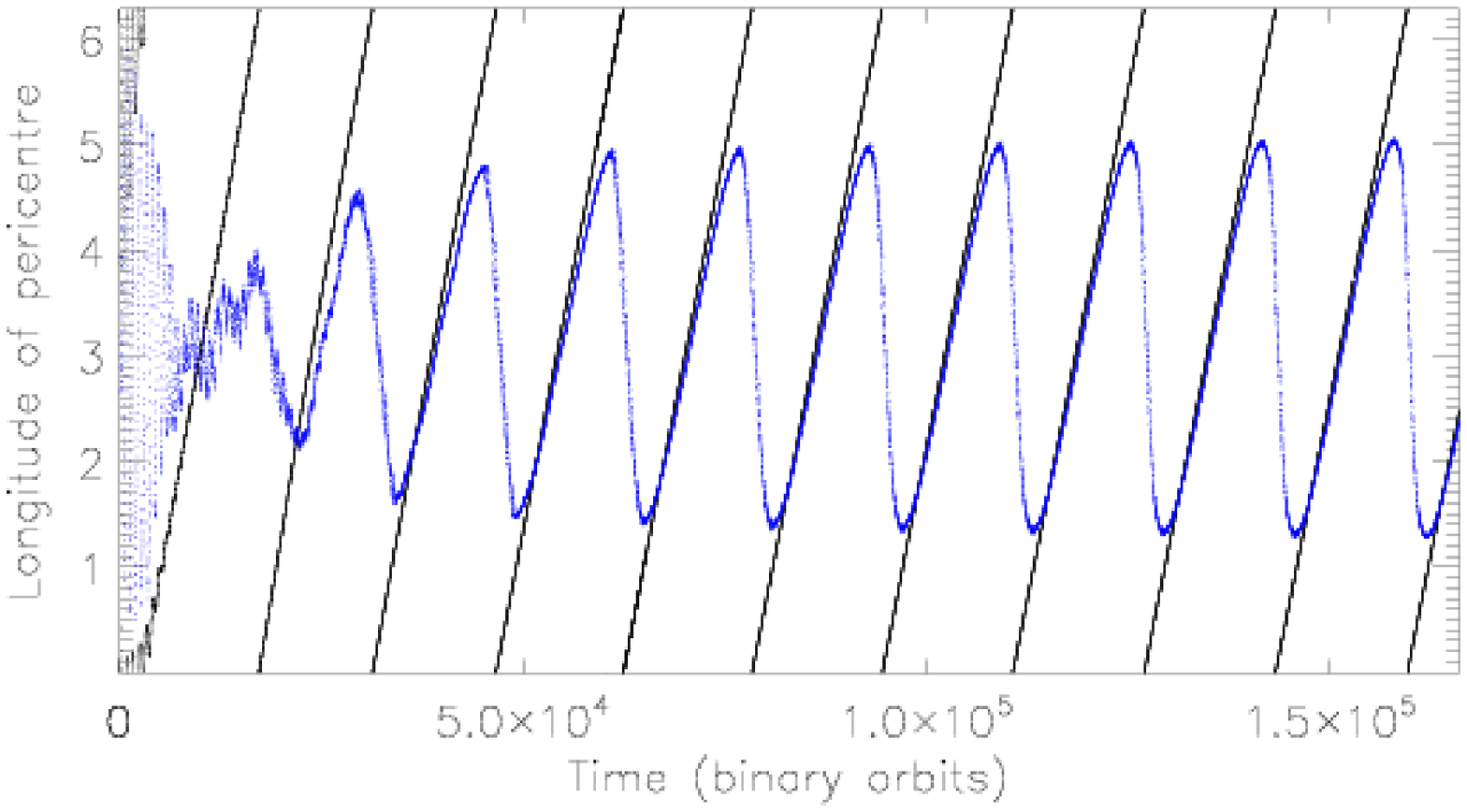}
   \includegraphics[width=\columnwidth]{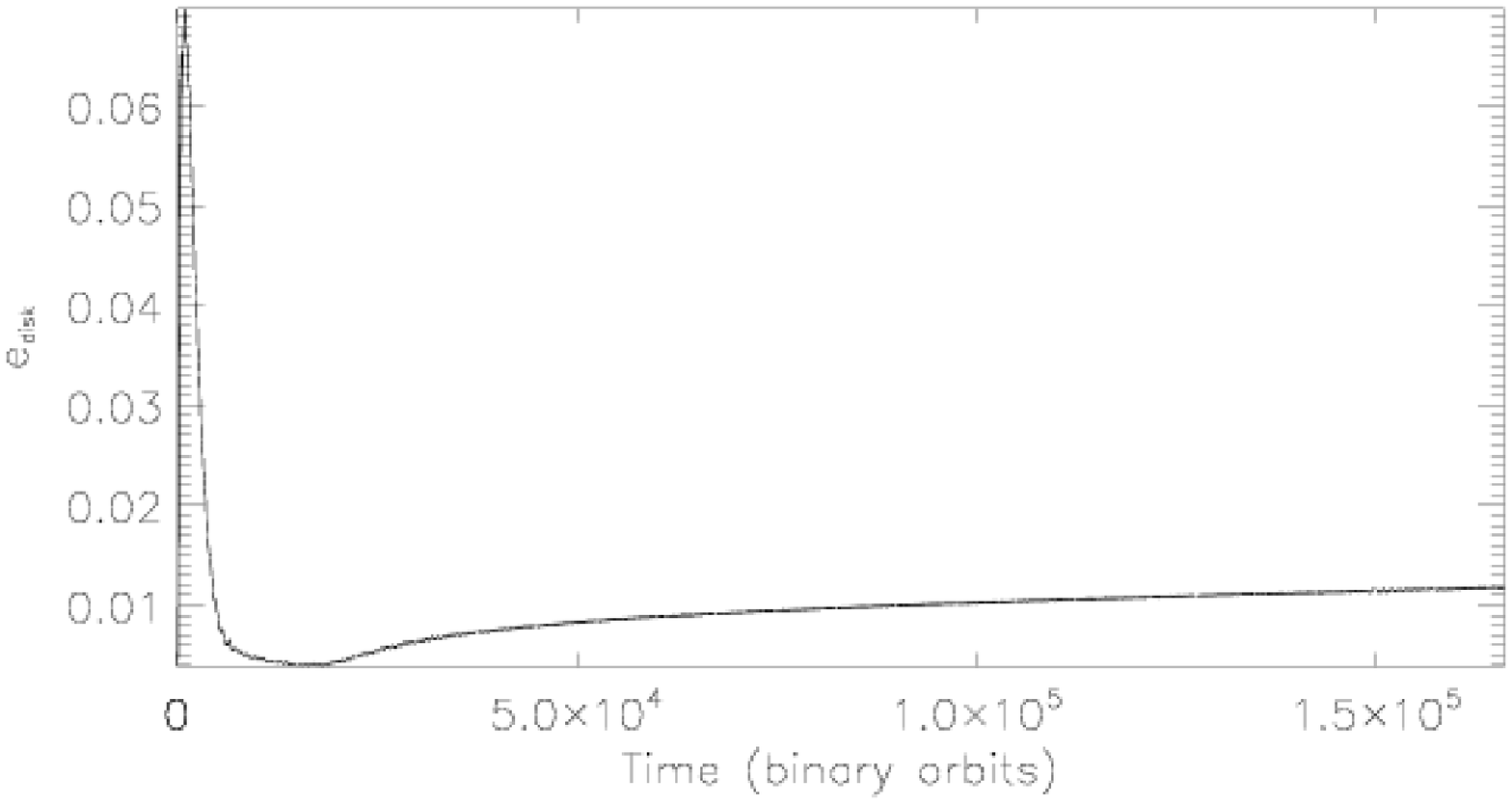}
      \caption{Same as Figure \ref{w_disk+bin}  but for 
high-resolution simulation with $256\times 380$ grid cells.
              }
         \label{w_disk+bin_hr}
   \end{figure}

Figures \ref{orbit_bin_hr} and \ref{w_disk+bin_hr} show the results of 
higher resolution simulations with  $256 \times 380$ grid cells. 
Figure~\ref{orbit_bin_hr} shows the evolution of the binary semimajor
axis and eccentricity, and fig.~\ref{w_disk+bin_hr} shows the
longitudes of pericentre of disk and binary (upper panel), and the
disk eccentricity (lower panel).
Because of their expensive
computing time, high-resolution simulations cannot be run for the same
length of time as low-resolution ones, and we have been unable to run
this simulation until the binary and disk eccentricities
have completely saturated. Nevertheless, we can see by 
comparing the results from the low and high resolutuion runs,
that the two simulations are in excellent agreement. At a time
of $\simeq 1.5 \times 10^5$ binary orbits, we see that binary 
eccentricity is $e_b \simeq 0.07$ in both cases, and the disk
eccentricity is $\simeq 0.01$ in each case.
We see from fig.~\ref{w_disk+bin_hr} that the apsidal lines of the disk 
and the binary are almost aligned at the end of the simulations,
such that it
appears that the system is close to an equilibrium state. It is
therefore reasonable to use the final result of this high-resolution run
as the initial condition for simulations dealing with the evolution of
protoplanets embedded in circumbinary disks. We present the 
results of such simulations in the following section. 

\subsection{Migration of planets in circumbinary disks.}
\begin{figure}
   \centering
   \includegraphics[width=\columnwidth,angle=0]{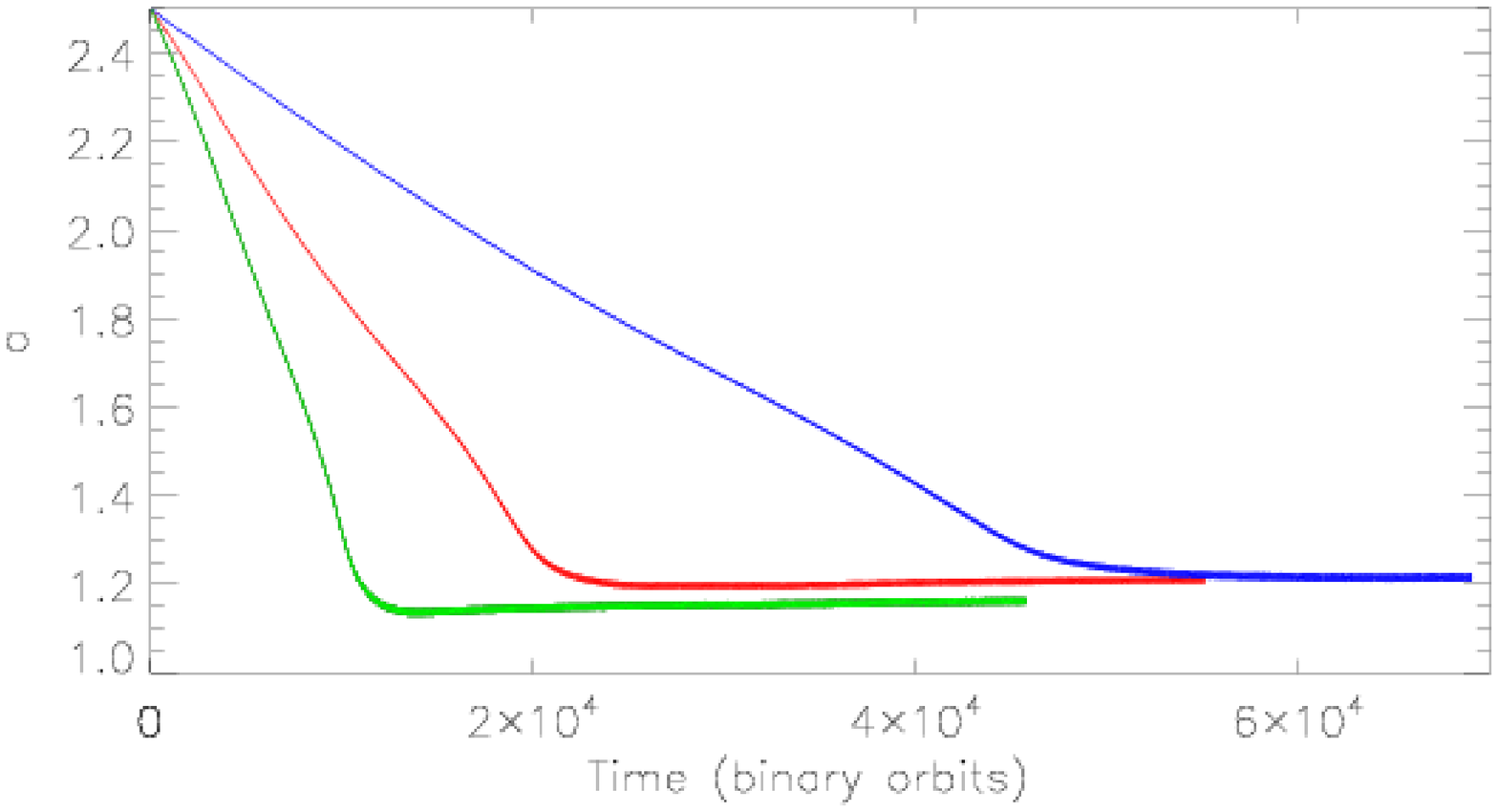}
    \includegraphics[width=\columnwidth,angle=0]{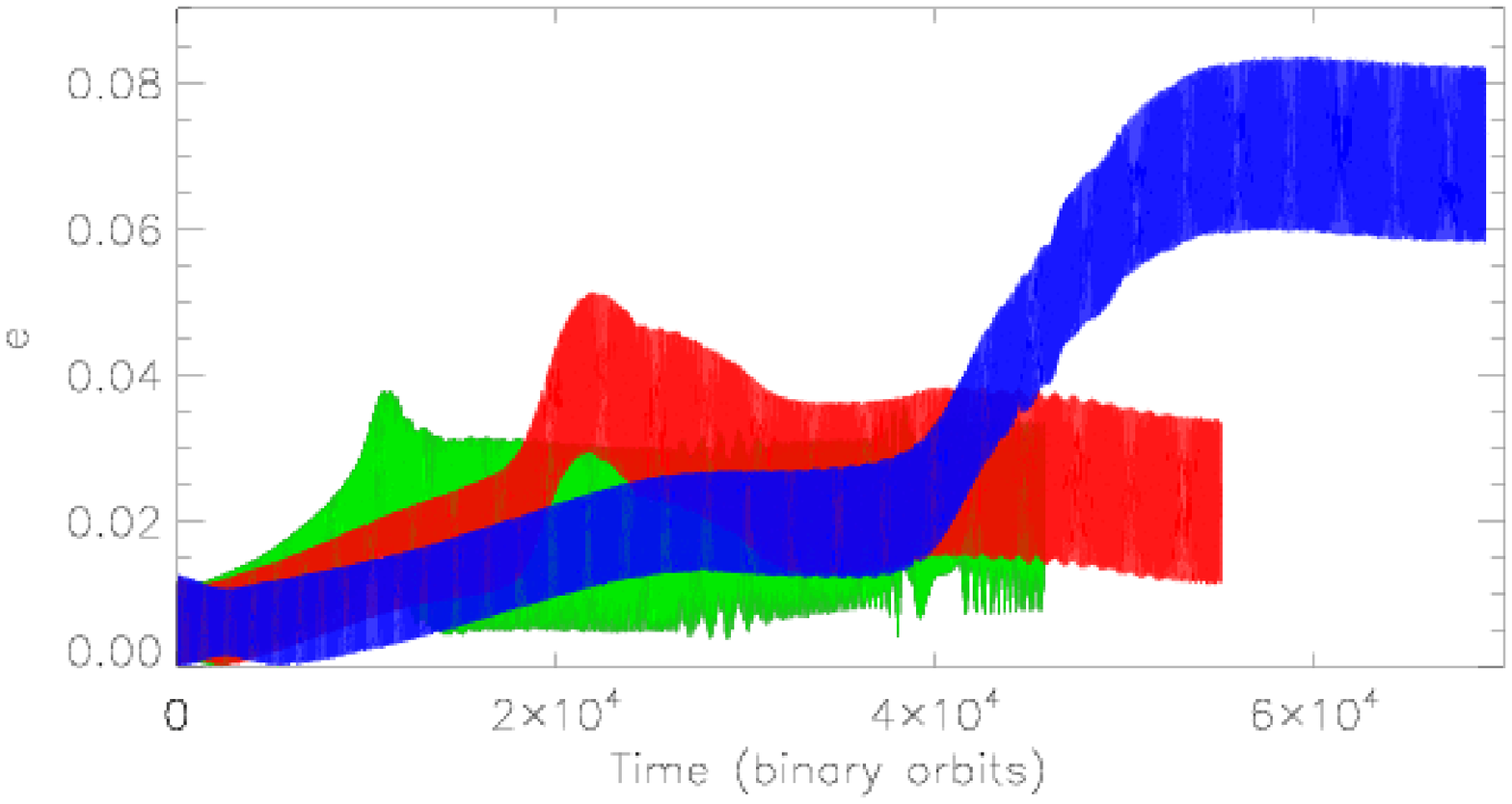}
    \caption{The upper panel of this figure shows the semi-major axis
              of the planets as a function of time, for three different
              planet masses. The lower panel displays the eccentricity 
              evolution of all three planets. }
         \label{orbit_pla}
   \end{figure}

We have performed three simulations with embedded protoplanets of mass
$m_p=5$, 10 and 20 $M_\oplus$, respectively. 
These masses are small enough that we expect the interaction between the 
disk and the planet to be linear, although the $m_p=20$ M$_{\oplus}$ case
is marginal. We insert these planets in the high resolution
circumbinary disk on circular orbit at $r=2.5$,
and let them evolve under the influence of both the disk and binary. 
The results are presented in Figure \ref{orbit_pla}, which shows the 
evolution of the planetary semi-major axis $a_p$ (top panel)
and eccentricity $e_p$ (bottom panel). 

As expected, each protoplanet initially undergoes a period of inward 
migration due to the torques exerted by the disk. We find migration 
rates consistent with analytical estimates for type I migration (e.g.
Tanaka, Takeuchi \& Ward 2002). At later times, all the 
protoplanets stop their migration at about the
same distance from the binary, located at $r\sim$ 1.1--1.2. We discuss
the reasons for this stalled migration below. 
As the planet migrates in, $e_p$ slowly increases because the damping 
of $e_p $ is not strong enough to counterbalance the eccentricity 
growth due to the interaction with binary. We see that the final eccentricity
is inversely proportional to the protoplanet mass, simply because the 
disk induced eccentricity damping rate is also inversely proportional to
the planet mass (e.g. Papaloizou \& Larwood 2000; Tanaka \& Ward 2004).

\begin{figure}
   \centering
   \includegraphics[width=\columnwidth,angle=0]{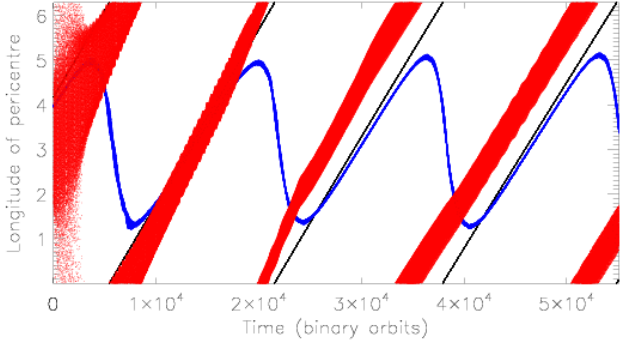}
    \caption{
              This shows the evolution of the longitudes of pericentre for the
              binary (black line), disk (blue line) and planet (red line).}
         \label{w-planet}
   \end{figure}

Figure~\ref{w-planet} shows the evolution of the longitudes of pericentre of
the binary, disk and protoplanet for the run with $m_p=5$ M$_{\oplus}$.
We see that the disk and binary are close to alignment, and precess
in a prograde direction at more of less the same rate.
The protoplanet eventually precesses at the same rate, but maintains a
constant angle of misaligment equal to about 0.5 radians.

\begin{figure}
   \centering
   \includegraphics[width=0.4\textwidth]{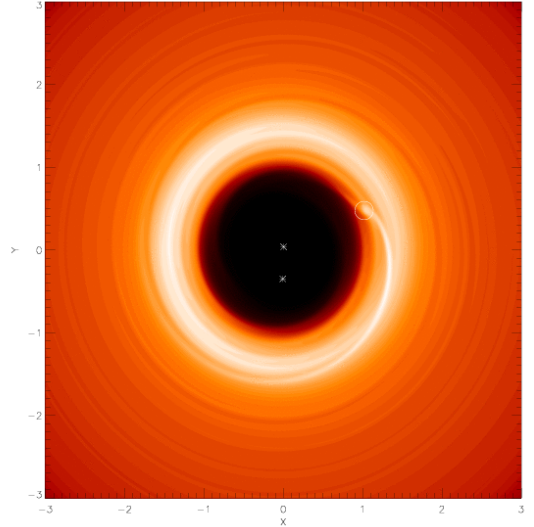}
    \includegraphics[width=0.4\textwidth]{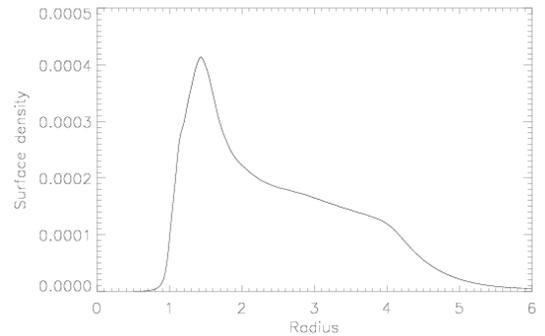}
      \caption{The upper panel of this figure shows a snapshot of disk
              surface density after $2 \times 10^4$ binary orbits. At this
              time, the migration of the planet is stopped. The lower
              panel shows the  azimuthal average of 
              the disk surface density at the same time.
              }
         \label{rho_2d_pla}
   \end{figure}

Figure~\ref{rho_2d_pla} shows a snapshot of the surface  density and the
corresponding azimuthal average after $2 \times 10^4$ binary orbits for
the simulation with $m_p=20$ M$_{\oplus}$, when the
planet has stopped its migration. 

In order to explain the stalling of migration, we have searched for
mean motion resonances between the protoplanets and binary system.
Indeed the semimajor axis $a_p \simeq 1.2$  at which the planets stall
is very close to the 5:1
mean motion resonance with the binary. We found no evidence, however,
that the protoplanets are in mean motion resonance.
Moreover, we find that migration stops even in test
simulations in which the planet does not feel the force due to the
secondary star. This is strong evidence that the effect we observe
is not due to the binary but instead arises because of the disk. \\

\begin{figure}
   \centering
  
   \includegraphics[width=\columnwidth]{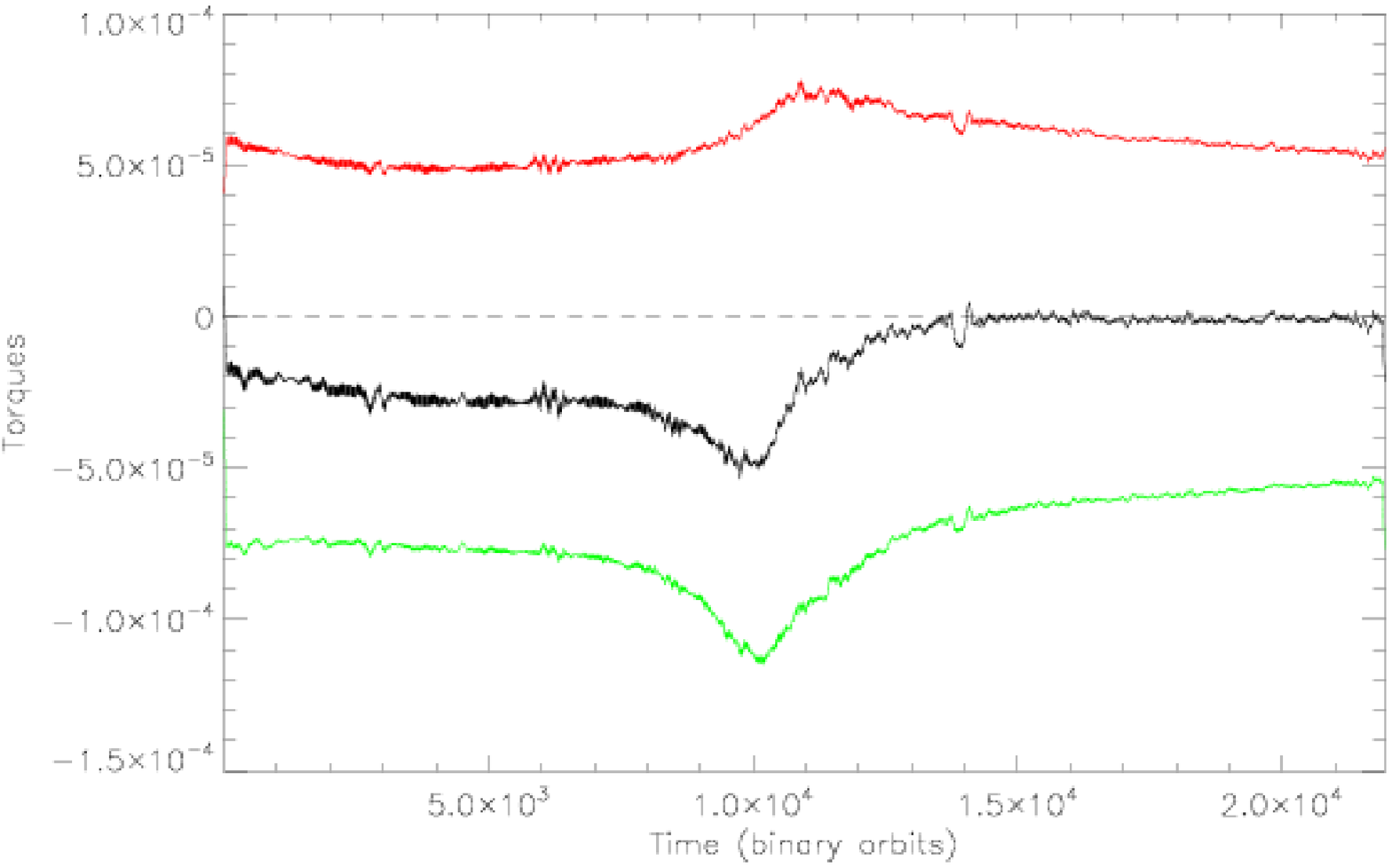}
   \includegraphics[width=\columnwidth]{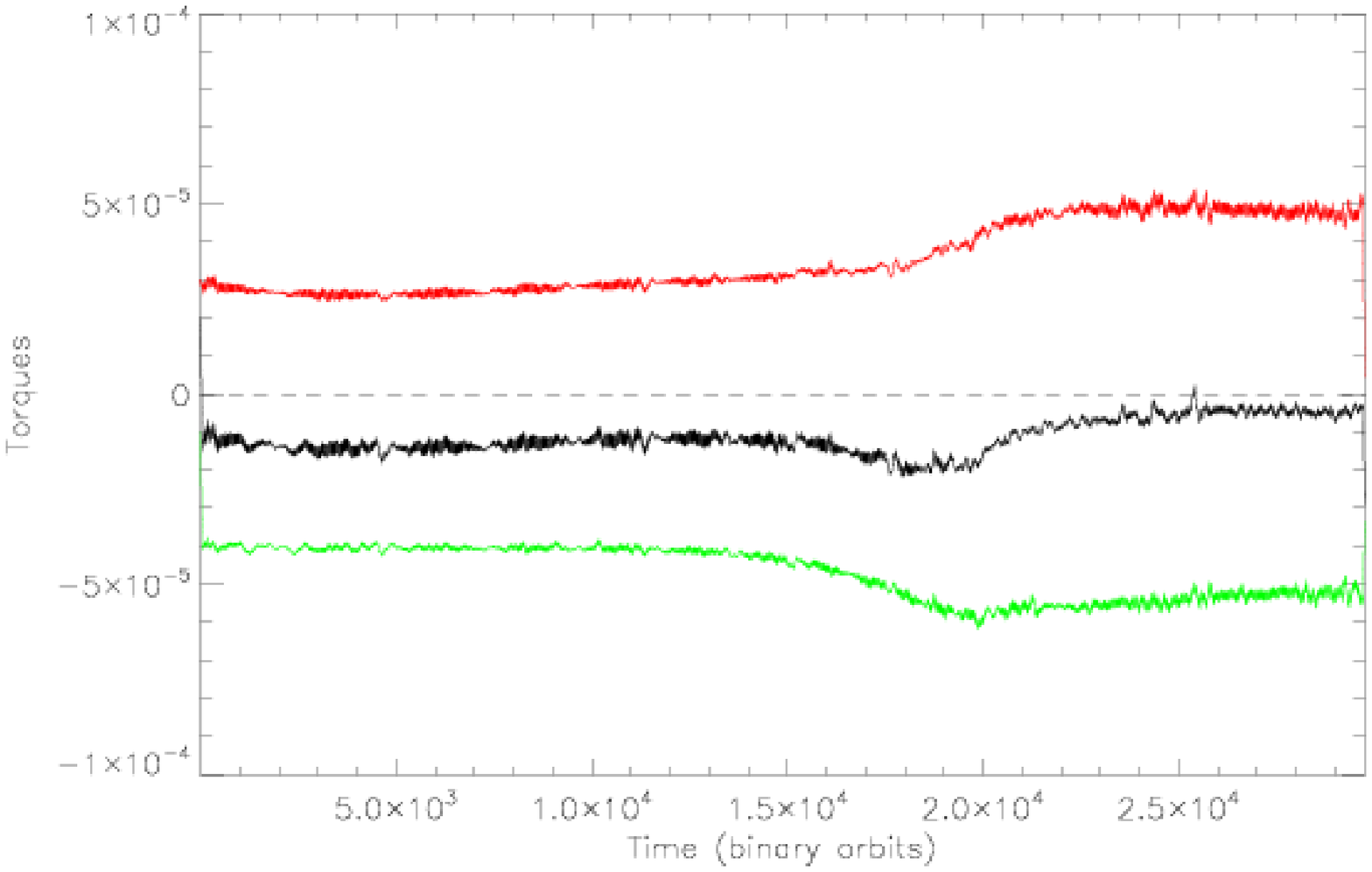}
      \caption{This figure shows respectively the torques exerted by the disk on a 20 $M_\oplus$ (upper
              panel) and a
              10 $M_\oplus$ (lowe panel) as a
              function of time. The
              green (resp. red)
              line  corresponds to the torques exerted by the inner
              (resp. outer) disk. The black line corresponds to the
              total torque. 
              }
         \label{torque}
   \end{figure}

Figure \ref{torque} shows the total torques exerted by the disk on the
planet as well as the torques due to the disk interior to $a_p$ and
exterior to $a_p$ for the simulations with $m_p=20$ and 10 M$_{\oplus}$.
As expected, the outer torques are initially larger
than the inner ones, corresponding to the inward migration of the
planet. The  10 M$_\oplus$ planet undergoes a period of
  accelerated migration at around $t=2\times 10^4$  binary orbits
  while the 20 M$_\oplus$ planet undergoes  accelerated migration at around $t=10^4$ binary orbits. This arises when the planets approach the tidally maintained inner cavity where the disk surface density is higher, resulting in an increased differential Lindblad torque acting on the planet. Because of the buffer action of the pressure gradient (Ward 1997), the strong density gradient at $r\sim 1.5$ (see lower
panel of Figure \ref{rho_2d_pla}) acts in the same way. Furthermore,
it leads to a strong negative corotation torque, which
promotes also an accelerated migration. After $\sim 2\times 10^4$ binary
orbits, the total torque exerted on the planet goes to zero, 
which is consistent
with the stopping of migration. The outer torques couterbalance
the inner ones exactly from this time onward.\\
The cancellation of the total torque appears to arise because 
the corotation torques increase in regions of strong positive 
surface density gradients, and can become strong enough to cancel out 
the differential Lindblad torque. This occurs when the planet enters the
inner cavity maintained by the binary tidal torques.
This effect was originally
discovered by Masset et al. (2006), who have
shown that in a disk with a surface density transition 
of about $50 \%$ occuring on a radial length scale equivalent to
3--5 disk vertical scale heights,
corotation torques (which are positive) can equal or exceed the
differential Lindblad torque (which is negative) near the
transition. They found that a protoplanet undergoing type I migration
can reach a fixed point at the transition where the total torque
cancels out and where planetary migration stops. This effect comes
into play provided that the length scale of the surface density jump
$\lambda$ is $\le 8HC/3$, where $H$ is the disk scale height at the
transition and  $C$ is a constant of order unity which depends on the profile of the density jump. In the simulations presented in this work, $\lambda \sim
2H$, which indicates that it is indeed the corotation torque which
is responsible for the observed stalling of migration near the gap edge.\\

\begin{figure}
   \centering
 \includegraphics[width=\columnwidth]{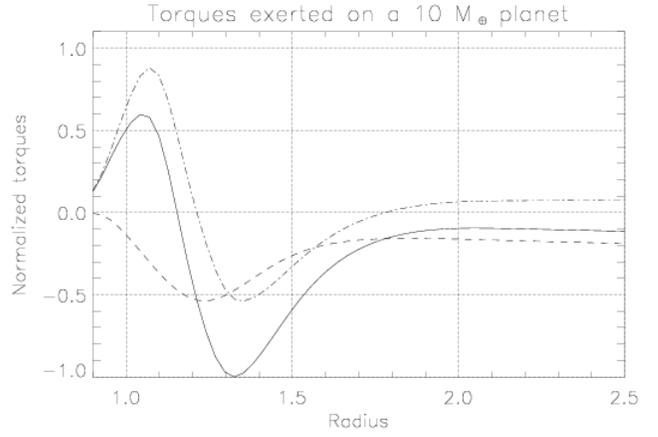} 
      \caption{This figure shows the semi-analytic torques exerted on a 
              10 $M_{\oplus}$ protoplanet as a
              function of the orbital radius. The dashed line
              represents the Lindblad torques and the dot-dashed line
              represents the corotation torques. The solid line shows
              the total (Lindblad + corotation) torques.
              }
         \label{analytic_torques}
   \end{figure}

In order to clearly demonstrate that this effect is at work here, we
have computed semi-analytically the corotation torques as well as the
Lindblad torques exerted on the planet. Corotation and Lindblad torques 
can be computed
in the linear regime from standard formulae 
(e.g. Goldreich \& Tremaine 1980) using
the surface density and rotation profile obtained in our simulations. Results
of these calculations for a 10 $M_{\oplus}$ planet are displayed in Figure
\ref{analytic_torques} which clearly shows that the corotation torques
can be larger that the Lindblad torques near the gap edge.
In agreement with Masset et al. (2006), we find also that the total
torque exhibits a fixed point at $r \sim 1.15$ where the corotation and 
Lindblad torques cancel each other. This value
corresponds approximately to the location where the planet stops its
migration in the hydrodynamical simulations. For $m_p=20M_\oplus$,
the location where the planet stops exhibits a small deviation with
respect to the other cases. This is probably due to the onset of
  non-linear effects which significantly alter the surface density
  profile and the corotation torque.

\subsection{Corotation torque saturation and turbulence}
The trapping of planets at the edge of the tidally truncated cavity
in circumbinary discs requires the action of corotation torques, and
it is well known that these can saturate in the absence of viscosity or
some other dissipative mechanism that is able to maintain the
surface density gradient in the vicinity of the planet
(e.g. Ogilvie \& Lubow 2003; Masset et al 2006).
It is worth noting that in a circumbinary disc the edge of the cavity is
maintained through a balance between tidal torques and viscous stresses,
such that even in a disc with very low viscosity it is possible that
saturation of the corotation torques may be prevented by
the tidal stresses due to the binary acting in the disc,
rather than through the viscosity itself.

In the simulations we present in this paper the viscous stress parameter
$\alpha=10^{-4}$, and this value is probably large enough to
prevent corotation torque saturation since it is consistent with the
minimum value required to prevent saturation derived in
Masset et al (2006). The origin of anomalous viscosity in accretion
discs is thought to be MHD turbulence generated {\em via} the
magnetorotational instability (MRI) (Balbus \& Hawley 1991).
Explicitly simulating a circumbinary disc that is undergoing MHD
turbulence is beyond current computational capabilities, but we note
that ideal MHD simulations of the MRI for zero net flux magnetic
fields typically yield values of $\alpha$ in the range
$10^{-3} - 10^{-2}$
(Brandeburg et al 2005; Hawley et al. 2006; Papaloizou \& Nelson 2003).
These values that are clearly large enough to prevent
saturation of corotation torques.
Questions remain about the level of MHD turbulence that can
be sustained in protostellar discs, since the ionisation fraction
is expected to be very low in significant regions of these discs
(Gammie 1996). The primary source of ionisation is likely to be
X--rays from the central protostar(s) (Glassgold et al 1997),
and chemical models of discs irradiated by X--rays indicate that
regions close to the star and the upper disc layers will be sufficiently
ionised to maintain MHD turbulence (Fromang et al 2002;
Ilgner \& Nelson 2006). In the case of a circumbinary case with
tidally truncated cavity, the cavity edge is face-on to the binary and
will therefore be exposed to a significant X--ray flux, so we should
expect that this region of the disc will be able to sustain
MHD turbulence that could in principle prevent  the saturation
of corotation torques.

It is worth noting that there are outstanding issues
about how corotation torques actually
operate in a turbulent disc. First, it has been shown that low
mass planets in turbulent discs experience stochastic forcing due to
turbulent density fluctuations (e.g. Nelson \& Papaloizou 2004;
Nelson 2005), and these may be strong enough to cause
the planet to `random walk' across the edge of the tidally truncated
cavity where the corotation torques operate to prevent inward
planet migration.
In addition, the operation of corotation torques for a 20 M$_{\oplus}$
protoplanet requires the fluid streamlines to form horseshoe--like orbits
in the corotation region, and it remains unclear whether these
horseshoe streamlines actually exist in a turbulent protoplanetary disc.
These issues are of relevance to the effectiveness of corotation torques
that operate in discs around single stars as well as in
circumbinary discs, and will be explored in future publications.

\section{Summary and conclusion}

In this paper, we have presented hydrodynamical simulations of low-mass planets
embedded in a circumbinary disk. The planets have masses ranging from
5 to 20 $M_\oplus$ and evolve under the forces exerted by the
disk and binary.\\
The central binary has mass ratio $q_{b}=0.1$. Values for the binary
eccentricity $e_{b}$ and semi-major axis $a_{b}$ are determined
self-consistently from simulations of an initially circular binary with
$a_{b}=0.4$ interacting with a circumbinary
disk. In agreement with expectations, a cavity is formed at the inner
edge of the disk because of the tidal torques due to the binary. 
$a_{b}$ continually decreases
while  $e_{b}$ increases with time. We find that the
final outcome of the interaction between the disk and the binary is a
configuration where the disk and binary eccentricity reach
a quasi--steady state.
This occurs when the disk and binary precess in a prograde sense
at the same rate with the
apsidal lines of the disk and binary being aligned.
In this quasi--steady state, $e_{b}\sim0.08$ and the inner edge of the disc
is truncated at $\sim 2.5a_{b}$, where $a_{b}\sim 0.4$.\\
The protoplanets we considered are inserted in the disk once this equilibrium
configuration has been approched. After a period of type I inward
migration, all the
protoplanets were found to stop their migration at about the same
location $r_p\sim 1.15-1.2$, which is close to the edge of the inner cavity.
Apparently, this behaviour arises because positive corotation torques cancel
the negative Lindblad torques at this point in the disk, in agreement
with the simulations of Masset et al. (2006) who considered
the migration of planets in disks with surface density transitions.
We examined the total torques exerted  by
the disk on the planet which were found to cancel 
as soon as the migration stalls.
In addition, we computed semi-analytical values for the corotation and
Lindblad torques. This analysis shows that, depending on the planet
mass, migration should stop at $r_p\sim 1.2$, which is in good
agreement with the results of the simulations.\\

These results have a number of interesting consequences for
planet formation in circumbinary disks.
According to Holman \&
Wiegert (1999), the critical semi-major
axis for dynamical stability 
of a  planet orbiting a binary with  $e_{b}\sim 0.08$ and
$q_{b}=0.1$ is $a_p\sim 2.3 a_{b}$. The planets in
our simulations stall at a distance of $a_p\sim 3 a_{b}$.
This suggests that migration
stops at a location where the planet should be stable long
after the dispersal of the circumbinary disk. From an observational
point of view, it means that the location corresponding to the
cavity edge of the precursor circumbinary disk appears to be an 
excellent place to look for low mass planets in close binary systems.\\
The final outcome of the system remains unclear, however,
if the protoplanet grows to become a giant planet. As the
protoplanet grows, gap formation can exclude the material from the
coorbital region, cancelling the effects of the corotation
torques. Once the gap is cleared, the planet should migrate inward
again on a time scale consistent with the type II regime (e.g. Lin \&
Papaloizou 1993; Nelson et al. 2000). 
Trapping into the 4:1 resonance
with the binary seems to be the most likely fate of a giant planet
interacting with a circumbinary disk (Nelson 2003). The evolution of
protoplanets which can accrete gas from the disk will be the subject
of a future publication. 

A number of other issues remain to be resolved, which will be
the subject of future papers. The first is how multiple low
mass protoplanets
will interact if they form at large distance from the binary and
successively migrate toward the cavity edge. It remains an open
question whether or not they will merge in this region to form
a larger body, become locked in mean motion resonances, or simply
undergo dynamical scattering. Also, the ability of planetesimals
to accrete in circumbinary disks needs to be examined, with particular
care being taken to simulate the structure of the circumbinary disk.
It seems unlikely that planets will form near the cavity edge,
but presumably there is a location outside of this which
remains benign to planetary formation. Once formed, these
planets will then migrate inward toward the binary
in the manner described in this paper.

\begin{acknowledgements}
The simulations performed in this paper were performed on
the QMUL High Performance Computing facility purchased
under the SRIF iniative, and on the U.K. Astrophysical Fluids Facility.
\end{acknowledgements}

\end{document}